\documentclass[%
 reprint,
nofootinbib,
amsmath,amssymb,
aps,
]{revtex4-2}

\usepackage{float}
\usepackage{graphicx}
\usepackage{dcolumn}
\usepackage{bm}
\usepackage{natbib}

\setcitestyle{authoryear,open={(},close={)}} 
\usepackage{url}
\usepackage{footnote}
\usepackage{graphicx}
\usepackage{txfonts}
\usepackage{tikz}
\usetikzlibrary{calc}
\usetikzlibrary{positioning}
\usetikzlibrary{shapes.geometric}

\usepackage{enumitem}

\usepackage[english]{babel} 

\def\degree{\hbox{$^\circ$\,}}

\def\gsim{\mathrel{\hbox{\rlap{\lower.55ex \hbox {$\sim$}}
                   \kern-.3em \raise.4ex \hbox{$>$}}}}
\def\lsim{\mathrel{\hbox{\rlap{\lower.55ex \hbox {$\sim$}}
                   \kern-.3em \raise.4ex \hbox{$<$}}}}

\def\he2{\hbox{He\,{\sc ii} $\lambda$4686}}

\def\RL1{\hbox{{$R_{L_{1}}$}}}

\newcommand{\aj}{Astron. J.} 
\newcommand{\aap}{Astron. Astrophys.} 
\newcommand{\aaps}{Astron. Astrophys. Suppl.} 
\newcommand{\mnras}{Mon. Not. R. Astron. Soc.} 
\newcommand{\pasp}{Publ. Astron. Soc. Pac.} 

\begin{document}


\title{\LARGE{Automated Detection of Satellite Trails in Ground-Based Observations Using U-Net and Hough Transform}}
\author{\large F.~Stoppa$^{1,2}$,
P.J.~Groot$^{1,3,4}$,
R.~Stuik$^5$,
P.~Vreeswijk$^1$,
S.~Bloemen$^1$,
D.L.A.~Pieterse$^1$,
P.A.~Woudt$^3$
}

\affiliation{
\begin{description}[labelsep=0.2em, align=right, labelwidth=0.7em, labelindent=0em, leftmargin=-2em, itemsep=0.5em, noitemsep]
  \raggedright
\item[$^{1}$] Department of Astrophysics/IMAPP, Radboud University, PO Box 9010, 6500 GL Nijmegen, The Netherlands
\item[$^{2}$] Department of Mathematics/IMAPP, Radboud University, PO Box 9010, 6500 GL Nijmegen, The Netherlands
\item[$^{3}$] Department of Astronomy and Inter-University Institute for Data Intensive Astronomy, University of Cape Town,  \\ 
\hspace{3em}Private Bag X3, Rondebosch, 7701, South Africa
\item[$^{4}$] South African Astronomical Observatory, P.O. Box 9, Observatory, 7935, South Africa
\item[$^{5}$] Leiden Observatory, Leiden University, Postbus 9513, 2300 RA Leiden, The Netherlands
\end{description}

}%

\date{\today\\}

\begin{abstract}

\noindent
\textit{Aims.} {The expansion of satellite constellations poses a significant challenge to optical ground-based astronomical observations, as satellite trails degrade observational data and compromise research quality. Addressing these challenges requires developing robust detection methods to enhance data processing pipelines, creating a reliable approach for detecting and analyzing satellite trails that can be easily reproduced and applied by other observatories and data processing groups.}

\noindent
\textit{Methods.} {Our method, called ASTA (Automated Satellite Tracking for Astronomy), combines deep learning and computer vision techniques for effective satellite trail detection. It employs a U-Net based deep learning network to initially detect trails, followed by a Probabilistic Hough Transform to refine the output.
ASTA's U-Net model was trained on a dataset with manually labelled full-field MeerLICHT images prepared using the LABKIT annotation tool, ensuring high-quality and precise annotations. This annotation process was crucial for the model to learn and generalize the characteristics of satellite trails effectively. Furthermore, the user-friendly LABKIT tool facilitated quick and efficient data refinements, streamlining the overall model development process.}

\noindent
\textit{Results.} {ASTA's performance was evaluated on a test set of 20,000 image patches, both with and without satellite trails, to rigorously assess its precision and recall. Additionally, ASTA was applied to approximately 200,000 full-field MeerLICHT images, demonstrating its effectiveness in identifying and characterizing satellite trails. The software's results were validated by cross-referencing detected trails with known public satellite catalogs, confirming its reliability and showcasing its ability to uncover previously untracked objects.}

\end{abstract}


\maketitle


\section{Introduction}

The rise of large satellite constellations has greatly enhanced global communication and connectivity. However, this development introduces significant challenges for ground-based telescopes, as satellite trails appear as streaks of light in astronomical images, degrading data quality by introducing noise and covering celestial objects \citep{McDowell2020}. Historically, astronomical observations have faced various sources of interference, but the rapid increase in satellite launches, especially with mega-constellations such as Starlink and OneWeb, has exacerbated the issue \citep{Tyson2020, Hainaut2020, Mallama2022, Bassa2022, Gallozzi2020, Groot2022}. Even space-based observatories, such as the Hubble Space Telescope, are not immune to these effects, highlighting the pervasive nature of satellite interference \citep{Kruk2023}.

The anticipated surge in satellite deployments could significantly increase the frequency and severity of interference, underscoring the need for proactive measures and innovative solutions to ensure the continued effectiveness of ground-based astronomy \citep{Walker2020}. Despite their advanced capabilities, wide-field optical ground-based telescopes, such as ATLAS \citep{Tonry18}, GOTO \citep{Steeghs22}, BlackGEM \citep{Groot2024}, ZTF \citep{Bellm19}, and Pan-STARRS \citep{Chambers16}, face significant challenges due to the increasing presence of satellite trails in their observations. Upcoming large aperture and wide field-of-view facilities like the Vera Rubin Observatory \citep{Rubin19} will be even more affected by these bright satellite trails. These streaks introduce noise and can obscure celestial objects, complicating the detection and analysis of transient events (see e.g., \citealt{Groot2022}).

Currently, several methods are employed to detect and mitigate satellite trails in astronomical images. These methods can be broadly categorized into simple source detection, template fitting to line shapes, computer vision techniques, and machine learning algorithms. Simple source detection, such as using SExtractor \citep{Bertin1996} and focusing on elongated shapes, has been employed to detect streaks. However, this approach is less effective at low signal-to-noise ratios (S/N) and tends to result in a high false-alarm rate \citep{Waszczak2017}. In contrast, template fitting to line shapes involves aligning a predefined streak shape to the image data and calculating a weighted sum of the pixels along the line to assess the match quality. This matched-filter approach \citep{Turin1960}, used by \citealt{Dawson2016}, employs the maximum likelihood method to detect streaks, assuming uncorrelated noise and a constant known Point Spread Function (PSF). While accurate, it can be computationally intensive and may require multiple templates or faster computational techniques.

\noindent
Computer vision techniques provide another set of tools for streak detection, using methods developed for natural image processing. These techniques are attractive due to their versatility and effectiveness in various imaging contexts. Two popular methods are the Hough Transform \citep{Duda1972} and the Radon Transform \citep{Radon1986}. The Hough Transform is applied to binary images following edge detection, while the Radon Transform is used on grayscale images. Despite their different applications, they are mathematically equivalent and have both been successful in finding streaks in crowded fields and images with diffuse light sources \citep{Cheselka1999, Virtanen2014, Bektešević2017}.
Both the Hough Transform and the Radon Transform have seen improvements in computational efficiency with the development of the Probabilistic Hough Transform and the Fast Radon Transform (FRT), respectively. These advanced methods, which often leverage GPUs, provide significant improvements in speed and computational performance \citep{Zimmer2013, Andersson2015, Nir2018, Borncamp2019}.

Despite recent progress, there remains a need for methods that can handle the increasing complexity and volume of astronomical data. Deep learning, in particular, offers new opportunities for improving streak detection. By training convolutional neural networks (CNNs, \citealp{LeCun1999}) on large datasets of labelled images, these methods can learn complex patterns and features directly from data, improving detection rates and reducing false positives \citep{Paillassa2020, Elhakiem2023, Chatterjee2024}.
Building on these advancements, our study introduces ASTA (Automated Satellite Tracking for Astronomy), a novel tool that uniquely combines the strengths of deep learning and computer vision techniques for detecting satellite trails in ground-based observations. We use a U-Net architecture \citep{Ronneberger2015} for initial satellite streak detection, followed by a Probabilistic Hough Transform to refine the output and extract satellite information.

A common criticism of machine learning methods is the difficulty in obtaining high-quality and well-labelled training sets, which are crucial for achieving accuracy and reliability. This issue often hinders the widespread adoption and reproducibility of such methods. 
However, we have addressed this challenge by using the LABKIT tool \citep{Arzt2022} to meticulously annotate images from the MeerLICHT telescope \citep{Bloemen2016}, ensuring the high-quality training data necessary for effective model training. This approach not only produces a reliable dataset for our application but also simplifies the creation process, making it accessible to other researchers and observatories. By providing a detailed account of our methodology and dataset preparation, we aim to facilitate the widespread adoption of ASTA, enhancing the ability of observatories worldwide to mitigate the impact of satellite trails on astronomical data quality.

The paper is organized as follows: Section \ref{sec: Data} describes the data and the process of image selection and preparation. Section \ref{sec: Method} outlines the machine learning techniques and the Probabilistic Hough Transform used for refining the detections, along with their validation. Section \ref{sec: Application} demonstrates the application of our tool, ASTA, by searching for Geostationary and Geosynchronous satellites in approximately 200,000 full-field MeerLICHT images. This analysis focuses on identifying satellite trails and cross-referencing the results with known public satellite catalogs. Finally, Section \ref{sec: Conclusions} summarizes our key findings and discusses the significance of developing sustainable solutions to ensure the advancement of astronomical research in the era of satellite constellations.

ASTA is directly accessible on GitHub \footnote{\url{https://github.com/FiorenSt/ASTA}}. All the images and masks used in this paper for training, test, and validation are available on Zenodo \footnote{\url{https://zenodo.org/records/11642424} } \citep{stoppa2024_data}.

\section{Data}
\label{sec: Data}

This study uses data from the MeerLICHT telescope, located in South Africa. MeerLICHT, a prototype for the recently operational BlackGEM telescope array \citep{Groot2024}, plays a crucial role in detecting and analyzing transient astronomical phenomena. Both MeerLICHT and BlackGEM are designed to capture high-quality astronomical data, but their ground-based nature makes them susceptible to satellite trails, which can degrade the quality of the data and impact the accurate analysis and interpretation of transient events.

In this section, we explain the data used to create ASTA and the dataset-building process. This involves selecting, preparing, and manually labelling images to ensure high-quality training data for our machine learning algorithm.

\subsection{MeerLICHT and BlackGEM Telescopes Array}

\begin{figure*}[ht]
    \centering
    \includegraphics[width=1\linewidth]{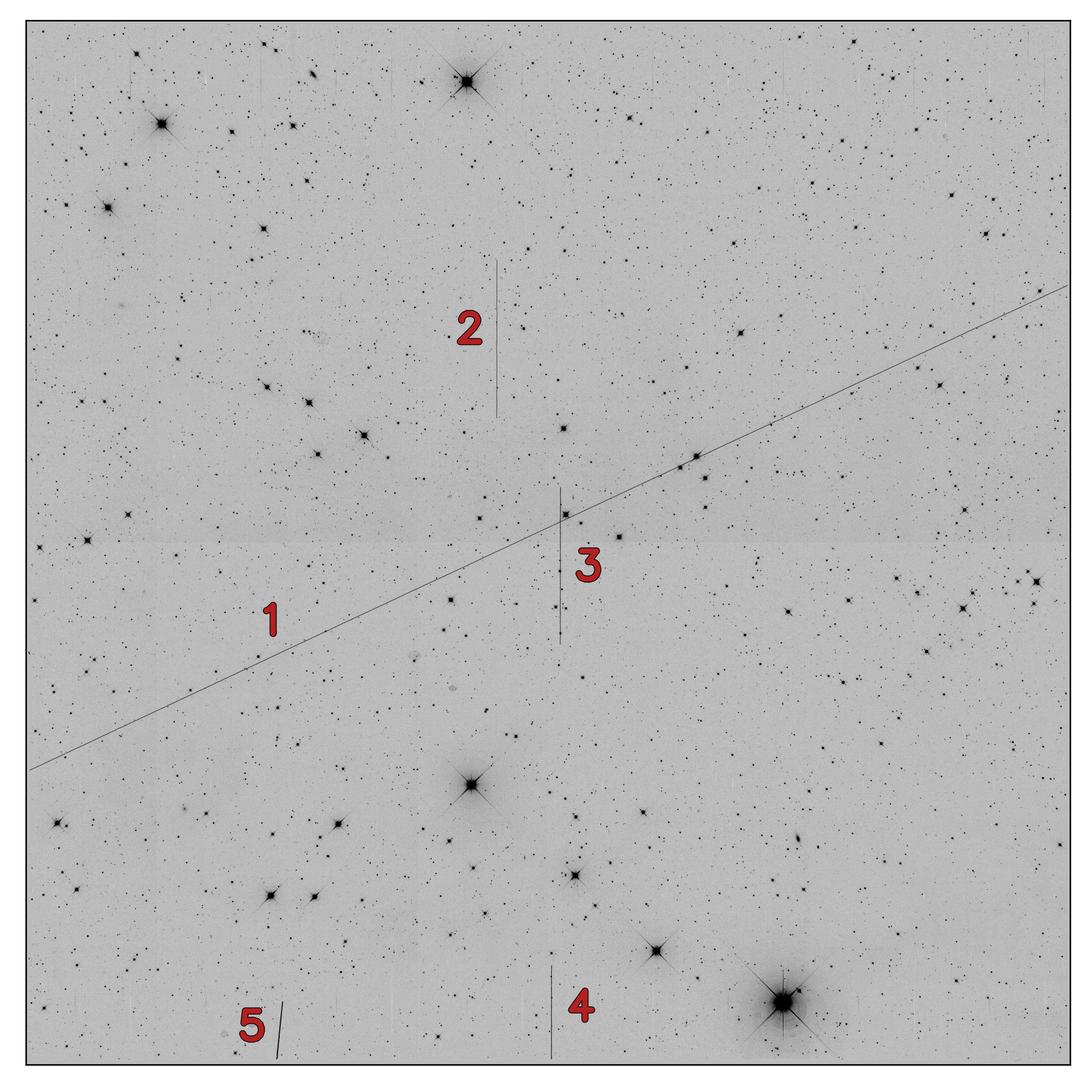}
    \caption{Full-field, $10560 \times 10560$ pixels, MeerLICHT image with a 60-second exposure, showcasing five satellite trails overlaying dozens of sources.}
    \label{fig: FullImageSet}
\end{figure*}

The MeerLICHT telescope, with its 65 cm aperture and a high-resolution $10.5k \times 10.5k$ pixel CCD, offers a wide field-of-view of 2.7 square degrees, sampled at $0.56^{\prime\prime}$/pixel.  MeerLICHT is equipped with the Sloan-Gunn type $u,g,r,i,z$ filter set, along with an additional wide-band $q$ filter (440-720 nm), enhancing its observational versatility. 
The telescope uses 60s integrations by default and reaches a limiting point-source magnitude of $q_{\text{AB}} > 20.5$ under standard conditions.
As a consequence, any satellite in Low Earth Orbit (LEO) up to Geostationary orbit (GEO) is very well detected and causes strong streaks in the observations. This causes interference in the photometric measurements of astrophysical objects but, at the same time, offers the opportunity to monitor the satellite's presence in Earth's orbit. 

Images captured by MeerLICHT are promptly processed at the IDIA/ilifu facility using BlackBOX \footnote{\url{https://github.com/pmvreeswijk/BlackBOX}} image processing software (Vreeswijk et al., in prep). The processing pipeline includes source detection via SourceExtractor \citep{Bertin1996}, astrometric and photometric calibration \citep{Lang2010}, PSF determination \citep{Bertin2011}, image subtraction, and transient detection \citep{ZOGY2016, Hosenie2021}. Furthermore, a new set of deep learning methods has been developed specifically for the MeerLICHT/BlackGEM telescopes. These methods are currently being tested and compared against traditional pixel-based counterpart techniques as detailed in \citealp{Stoppa2022, Stoppa2023_a, Stoppa2023_b}.

\noindent
Figure \ref{fig: FullImageSet} illustrates a typical full-field MeerLICHT image, showcasing the impact of multiple satellite trails.

\subsection{Dataset preparation}
\label{sec: Training Dataset}

\begin{figure}[ht]
\centering
\includegraphics[width=1\linewidth]{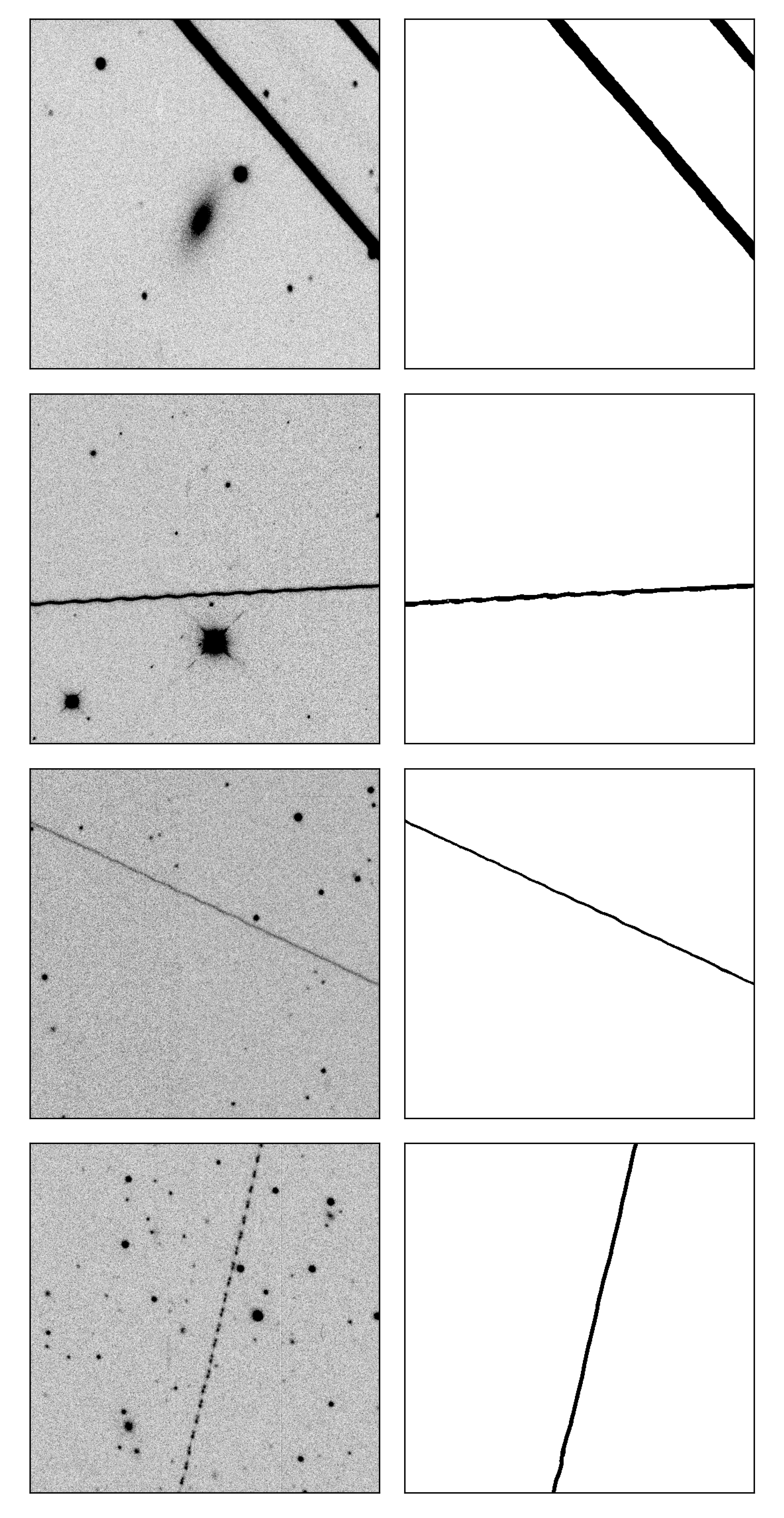}
\caption{Examples of satellite trail annotations in MeerLICHT images. The figure shows two columns and four rows: (a) Original images with satellite trails of varying intensities and types in the first column, (b) Corresponding labelled ground truth masks in the second column.}
\label{fig: Masks}
\end{figure}

Creating a high-quality training dataset is fundamental for developing an effective machine learning model. In our study, we require an accurately labelled dataset where pixels belonging to a satellite trail are assigned to one class, and all other pixels—including those representing the sky background, stars, and common linear artefacts seen in CCD images, such as diffraction spikes, cosmic ray hits, and charge bleeding—are assigned to another class. With such a dataset, the network can effectively learn the relationship between the original image and the segmentation mask, enabling accurate and reliable detection of satellite trails.

\noindent
However, a significant challenge for researchers attempting to reproduce machine learning methods for their specific applications is the lack of a well-labelled dataset and uncertainty about how to create one from scratch. This obstacle often discourages the adoption of machine learning techniques. To address this issue, we provide a detailed explanation of how to build a reliable dataset for satellite trail detection, offering a clear guide for other researchers and observatories.

We started by collecting 178 full-field MeerLICHT images, all visually inspected to identify the presence of satellite trails. These images included trails of varying lengths and signal-to-noise ratios, ensuring a diverse and comprehensive dataset. As can be seen in Fig. \ref{fig: Masks}, MeerLICHT's spatial resolution is high enough that seeing- and tracking-variations during the integration can be seen as a ‘wavy’ nature of the trail, making them significantly deviate from just a straight line.

To facilitate the creation of the ground truth segmentation masks, we used LABKIT \citep{Arzt2022}, a plugin for the Fiji image processing package, which simplifies the annotation process with its pixel classification algorithm for quick automatic segmentation.

\noindent
In LABKIT, we first manually classified a really small subset of pixels in the images ($\sim0.001\%$) into two classes: satellite trails and background. This initial step involved selecting pixels that represent the trails and marking the sky background, sources, and other linear features as background. LABKIT then used these labelled pixels and a set of features automatically extracted from the images to train a random forest classifier. The feature vector for each labelled pixel is obtained by applying a set of filters to the image, such as Gaussian, a difference of Gaussians, or Laplacian filters. These filters emphasize different features of the input image, and their responses for each pixel are added to the feature vector. The final feature vectors, paired with their respective ground-truth classes, constitute the training set for the random forest, which consists of a hundred decision trees. LABKIT then proceeds to predict the segmentation mask for the entire image and provides a quick, although rough, approximation of all the trails.

\noindent
After LABKIT's initial automated classification, the software easily provides tools to manually refine the annotations to ensure their accuracy and reliability. Using LABKIT’s interface, we adjusted the pixels and specific parts of trails or spurious detections that were assigned to the wrong class. 
This comprehensive annotation process resulted in a full binary mask for each full-field MeerLICHT image and was completed by a single person in three days.

After labelling, the full-field images and their associated ground truth masks were divided into smaller patches of 528x528 pixels to facilitate efficient training. Data augmentation techniques, including 90-degree rotations, flips, and shifts, were applied exclusively to patches containing satellite trails. This approach enhanced the model’s exposure to diverse trail characteristics, thereby improving detection accuracy. This selective augmentation was necessary to prevent the dataset from becoming unbalanced, as most patches from a full-field image do not contain satellite trails. Without this, the dataset would be dominated by patches with empty masks, making the model training more difficult and less effective.

\noindent
Figure \ref{fig: Masks} shows four examples of patches with satellite trails and their corresponding ground truth masks.

\section{Method}
\label{sec: Method}

This section outlines the methodology employed by ASTA for detecting and analyzing satellite trails in astronomical images. ASTA leverages a U-Net architecture for the initial segmentation of satellite trails, providing a robust framework for distinguishing trails from other features. To enhance the precision of trail delineation, the initial segmentation is refined using the Probabilistic Hough Transform \citep{Galamhos1999}. This combination of deep learning and classical image processing techniques ensures high accuracy and reliability in identifying and characterizing satellite trails.

\subsection{Detection of Satellite Trails Using U-Net}
\label{sec: UNET}

\begin{figure}[ht]
\centering
\includegraphics[width=1\linewidth]{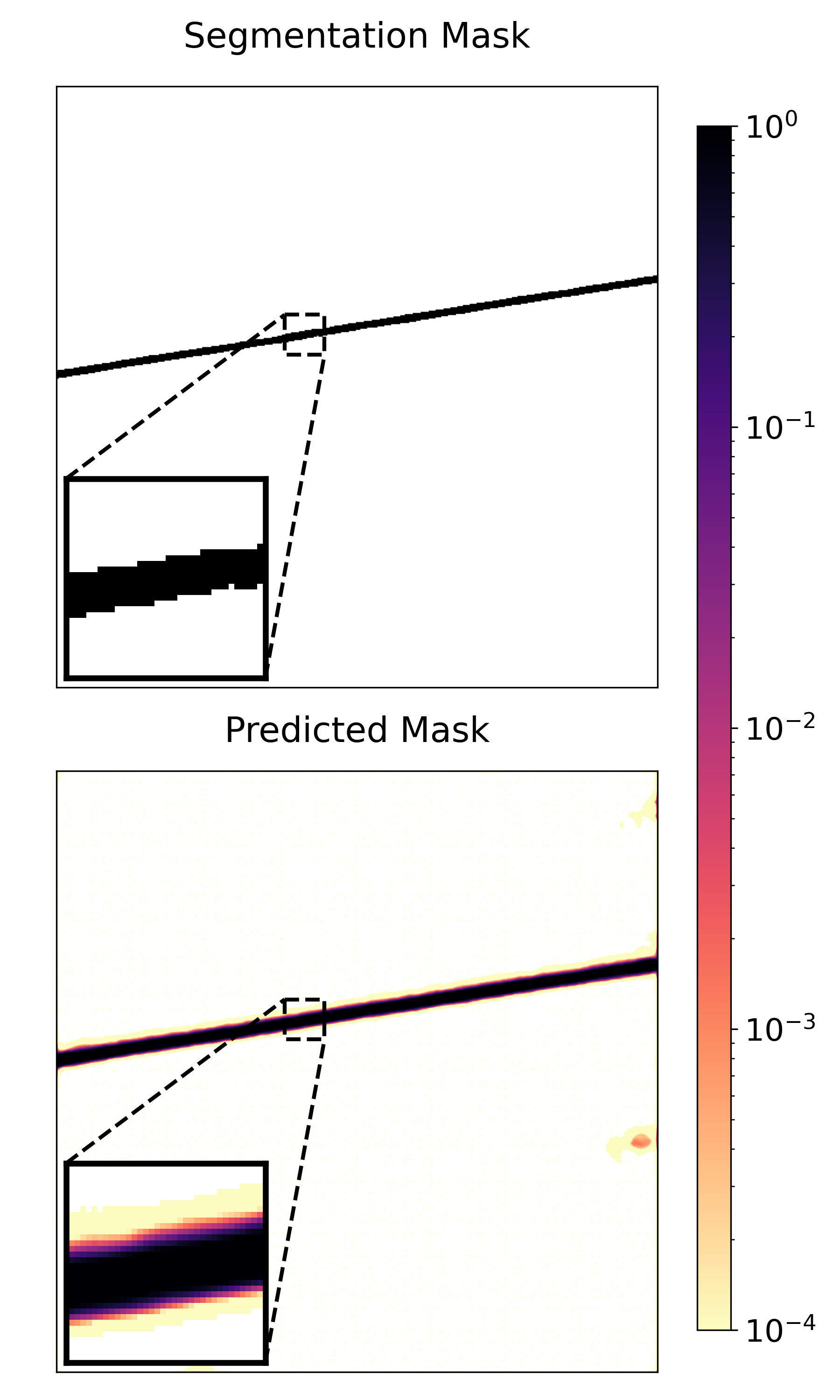}
\caption{Comparison of ground truth segmentation mask and U-Net predicted segmentation map. The top panel shows the ground truth segmentation mask with satellite trails marked in black. The bottom panel shows the U-Net predicted segmentation map, with pixel values ranging from 0 to 1, indicating the likelihood of each pixel belonging to a satellite trail. The insets provide a zoomed-in view to highlight the detailed accuracy of the predictions.}
\label{fig: UNET_prediction}
\end{figure}

Astronomical images are often crowded with various linear and non-linear features, including stars, galaxies, cosmic rays, and diffraction spikes from bright stars. These complexities present a significant challenge for accurately identifying satellite trails. The U-Net architecture \citep{Ronneberger2015}, originally developed for biomedical image segmentation, excels at identifying complex patterns in images, making it well-suited for detecting satellite trails against such diverse backgrounds.

\noindent
U-Net's architecture is designed to understand and reconstruct the context of an image through two main pathways: a contracting path that compresses the image to grasp its broader context, and an expansive path that reconstructs the image's details for precise localization of features. This structure allows U-Net to process the image at multiple scales, capturing both the overall patterns and the fine details. By integrating features from both paths, U-Net maintains a balance between contextual understanding and detailed segmentation.

Our network consists of convolutional layers with LeakyReLU activations that adjust the spatial dimensions and number of filters, from 8 to 128. To prevent overfitting, dropout layers are integrated within the network. The final layer produces a predicted segmentation map, indicating the likelihood of each pixel belonging to a satellite trail with values ranging from 0 to 1. Overall, the model has approximately 485,000 trainable parameters, ensuring it is both lightweight and efficient for large astronomical datasets.

\noindent
To optimize the model during training, we used the Combo loss \citep{Taghanaki2018}, a combination of binary cross-entropy (BCE) loss \citep{Mannor2005} and Dice loss \citep{Sudre2017}. This approach balances pixel-wise accuracy with segmentation performance, improving detection accuracy in class-imbalanced datasets. For a more detailed description of this loss function, we refer readers to \citealt{Stoppa2022}.

Figure \ref{fig: UNET_prediction} illustrates a comparison between a ground truth segmentation mask and the corresponding prediction made by the U-Net model. The predicted segmentation map effectively captures the trail, and the pixel values decrease rapidly to zero beyond the edges of the trail, indicating the U-Net's ability to accurately identify and differentiate the trail from the background and other artefacts.

After the U-Net processes the images, we apply a threshold to its output to create binary segmentation masks. This step is crucial as we need to identify which pixels belong to satellite trails and which do not. Pixels with values above the threshold are classified as satellite trails (value of 1), while those below are classified as background, including the sky and astronomical sources (value of 0). 

\begin{figure}[ht]
\centering
\includegraphics[width=1\linewidth]{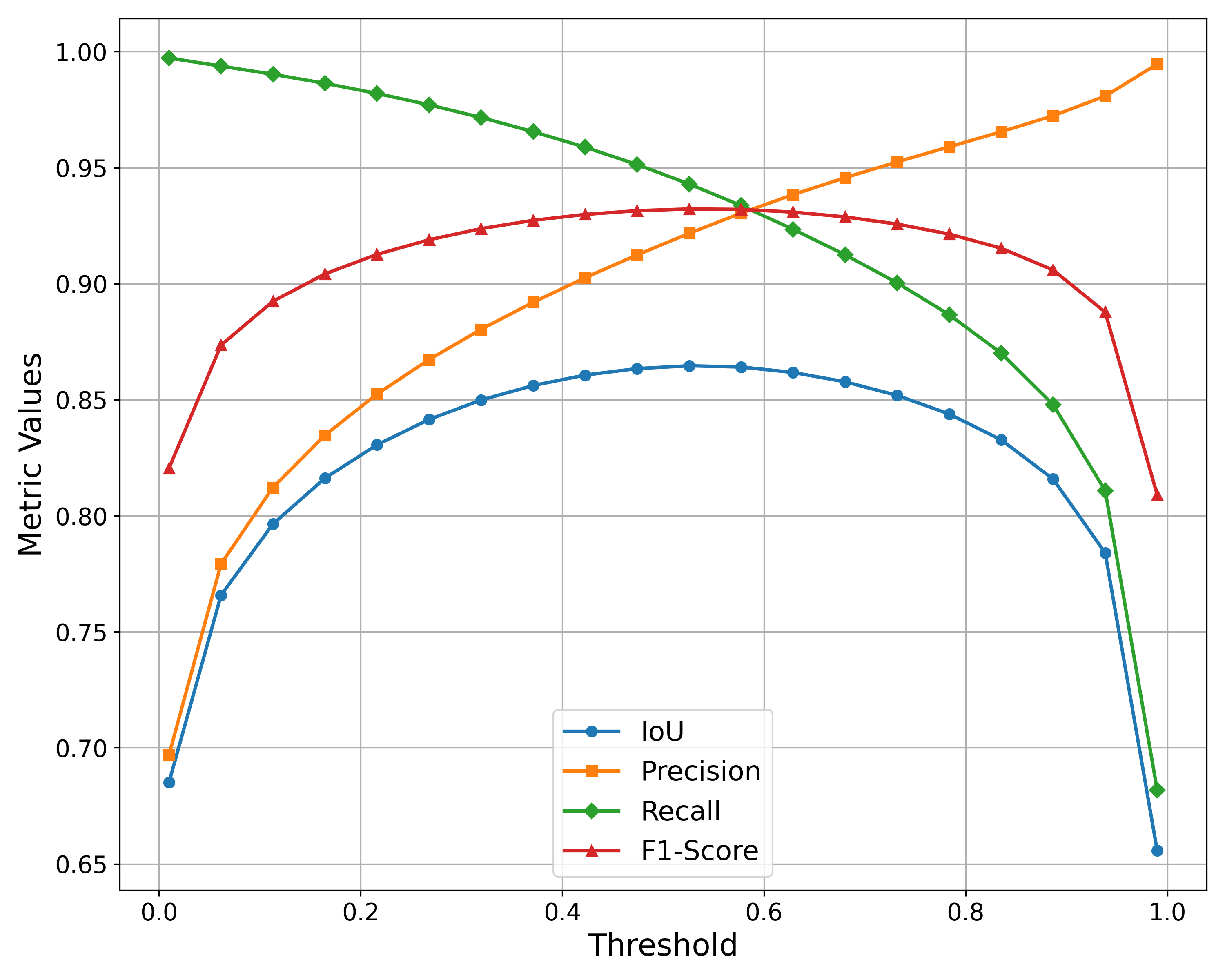}
\caption{Performance metrics for U-Net across different threshold levels: IoU, Precision, Recall, and F1-score. A threshold of 0.58 provides a balanced result in terms of all metrics tested.}
\label{fig: metrics}
\end{figure}

\noindent
To evaluate the effectiveness of the U-Net and determine the optimal threshold, we tested the model on 20,000 patches, both with and without satellite trails. We used several metrics, including precision, recall, F1-score, and Intersection over Union (IoU), to assess the model's performance at various threshold levels. Figure \ref{fig: metrics} shows these metrics across different thresholds.

\noindent
Precision measures the accuracy of the model's positive predictions, while recall assesses its ability to identify all relevant instances. The F1-score balances precision and recall, providing a single metric that accounts for both false positives and false negatives. Intersection over Union (IoU) measures the overlap between the predicted segmentation and the ground truth, offering a comprehensive view of segmentation quality.
Independently of the threshold, all metrics indicate that U-Net performs well in predicting trails and distinguishing them from other linear artefacts. A threshold of 0.58 provides a balanced result and is therefore used as the default value for the successive analyses in this paper.

\subsection{Refinement with Probabilistic Hough Transform}
\label{sec: Hough Transform}

\begin{figure*}[ht]
    \centering
    \includegraphics[width=1\linewidth]{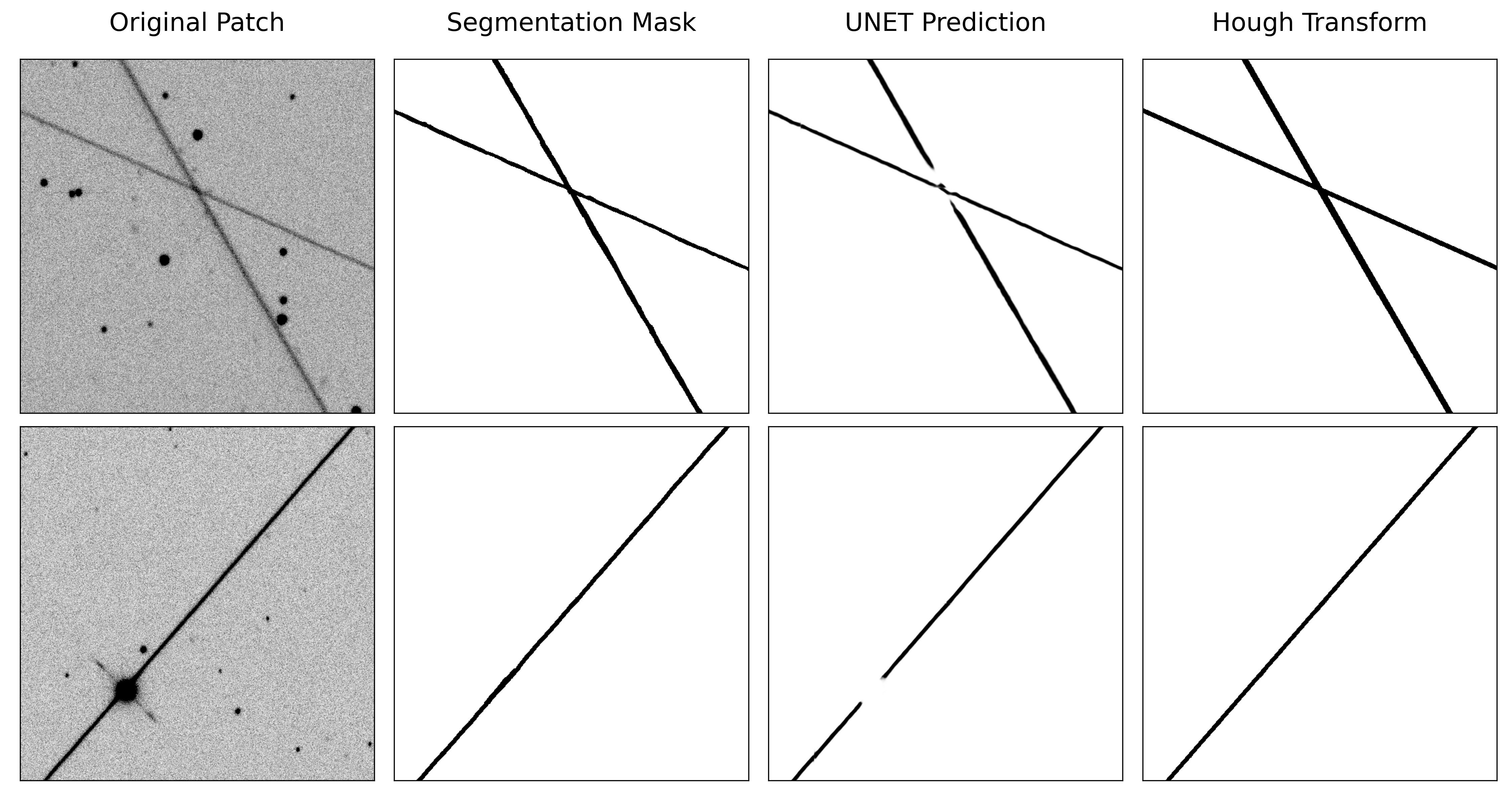}
    \caption{Sequential steps of satellite trail detection and refinement: (a) Original patch image, (b) Ground truth segmentation mask, (c) U-Net predicted segmentation map, (d) Final result after applying the Probabilistic Hough Transform. This workflow demonstrates the process from initial detection to refined trail delineation, ensuring precise identification and continuity of satellite trails.}
    \label{fig:steps}
\end{figure*}

While U-Net predictions effectively identify satellite trails, they may exhibit gaps due to factors such as tumbling of the satellites, bright stars, detector defects, or unfortunate locations in the $528 \times 528$ pixel patches. To address these gaps in the recombined full-field binary masks, we applied the Probabilistic Hough Transform \citep{Galamhos1999}, which is highly effective for detecting linear patterns in images. The primary function of the Hough Transform in this context is to fill in splits or gaps in the predicted trails, ensuring continuous and accurate representation. This refinement step maintains the integrity of trail detection, particularly in areas with discontinuities, and ensures more accurate statistics about satellites. This is crucial for further steps, such as estimating the total number of satellites and matching them with known satellite catalogues.

\noindent
The Hough Transform works by translating spatial relationships within an image into a parameter space, making it effective for detecting linear patterns like satellite trails. In this space, any line in the image can be represented as a point defined by the equation \(r = x\cos(\theta) + y\sin(\theta)\), where \(r\) is the perpendicular distance from the origin to the line, and \(\theta\) is the angle of this perpendicular line with the horizontal axis. For each pixel that might belong to a satellite trail, the Hough Transform evaluates every possible line through that pixel, represented by various (\(r\), \(\theta\)) combinations, resulting in a sinusoidal curve in the parameter space for each pixel. The intersection of these curves from different points indicates a consensus on the presence of a line in the image space, with accumulations in an array highlighting the most significant lines. A threshold is applied to distinguish meaningful lines from noise, ensuring that only genuine satellite trails are identified. 

\noindent
The Probabilistic Hough Transform, a variant of the standard Hough Transform, improves computational efficiency by considering a subset of points in the image to detect lines. Optimizing the resolution of the parameter space, particularly the granularity of the (\(r\), \(\theta\)) bins, is essential for accurately detecting satellite trails, balancing sensitivity and computational efficiency. This adjustment allows for precise identification of satellite trails, enhancing the reliability of our analysis.
Figure \ref{fig:steps} illustrates two cases where the U-Net prediction is improved by the Hough Transform, demonstrating the process from initial detection to refined trail delineation and ensuring precise identification and continuity of satellite trails.

\subsection{Contour Analysis and Feature Extraction}
\label{sec: Contour Analysis}

Following the refinement by the Hough Transform, we obtain a binary mask consisting solely of satellite trails. The final step involves extracting features of the detected trails, such as length, location of start and end points, inclination, and brightness.

\noindent
To achieve this, we identify the contours of the trails in the refined binary mask using the cv2.findContours function from the OpenCV package. If each satellite trail were independent, we would only need to extract the pixel values within the contour to determine the trail brightness and easily identify the most extreme points of the trail and their coordinates. However, the recent increase in satellite trails often results in multiple trails crossing each other, as shown in the first row of Fig \ref{fig:steps}.

\noindent
To address this occurrence, for each independent contour identified, there is an effective method to determine if the contour is actually the intersection of two or more trails. This is achieved by running a clustering algorithm based on DBSCAN (Density-Based Spatial Clustering of Applications with Noise, \citealp{Martin1996}) on the angles of the Hough Transform segments that compose the current contour. This can quickly identify as many clusters as there are intersecting trails and provides an effective solution for separating them. Once the trails are separated, features such as length, location of start and end points, inclination, and brightness are easily extracted.

\section{Application to MeerLICHT Data}
\label{sec: Application}

\begin{figure*}[ht]
    \centering
    \includegraphics[width=1\linewidth]{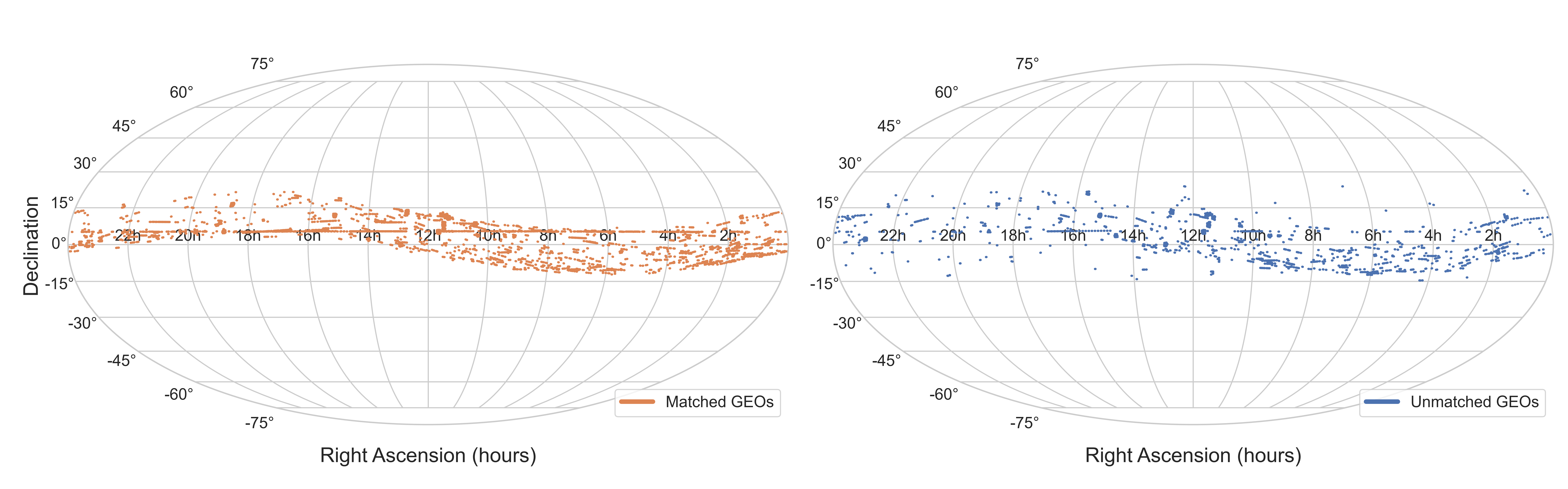}
    \caption{Matched (left) and unmatched (right) GEO/GES satellite trails in celestial equatorial coordinates. The geostationary belt is visible at Dec=+6$\degree$, as observed from South Africa. The sinusoidal band are the geosynchronous satellites. The unmatched trails show satellites in both the geostationary belt and in geosynchronous orbit, as well as more extreme cases at slightly higher and lower declinations.}
    \label{fig:GEO_Detection}
\end{figure*}

In this section, we apply ASTA to MeerLICHT images collected since January 2020. We analyzed approximately 200,000 non-red-flagged (i.e., science-grade) images, detecting both Low Earth Orbit (LEO) and Geostationary/Geosynchronous (GEO/GES) satellites. LEO satellites typically cause streaks from edge to edge of a full-field MeerLICHT image, as illustrated in Fig. \ref{fig: FullImageSet}, complicating their identification. These will be highlighted in a forthcoming paper. Here, we focus on the detection and identification of both GEO and GES satellites, whose trails generally start and stop within one exposure with a 60-second integration time.

GEO satellites orbit Earth at an altitude of approximately 35,786 kilometers, matching the planet’s rotational period. This allows them to remain stationary relative to a fixed point on Earth. GEOs are widely used for communication, weather monitoring, and broadcasting. In astronomical images, GEOs typically appear as short streaks with a length consistent with the integration time in seconds due to the fact that the telescope tracks celestial objects and not satellites. For MeerLICHT/BlackGEM, which use 60~s integration times, this means a streak of 15 arcminutes in length, corresponding to $\sim$1600 pixels, in the East-West direction (90$\degree$ in the convention used here). 

\noindent
GES satellites, on the other hand, have orbits with the same period as the Earth's rotation but are inclined relative to the equator. This results in their position in the sky tracing an analemma over time. These satellites can appear as longer or shorter trails depending on their current position in their orbit relative to the observer and are generally at an angle to the East-West orientation on the detector. 

While ASTA effectively detects satellite trails, some artefacts and image contaminations can mimic satellite signatures. Diffraction spikes, which are linear artefacts caused by bright stars, always align at 45$\degree$ and can be mistaken for satellite trails. Additionally, CCD defects such as dead or hot pixels and reflections from the Moon and/or local light sources such as car headlights can produce spurious linear patterns.
Specifically for MeerLICHT/BlackGEM, there is a class of reflections from bright stars just outside the field of view that appear as short horizontal (North-South) streaks on the edge of the detector due to an unwanted reflection in the cryostat. 
However, for this application to GEOs and GES satellites, these artefacts are less problematic since GEO trails are almost exactly in the East-West direction and geosynchronous satellites are centred at $90\degree$ with an inclination of $\pm 15\degree$.

Starting from ASTA's results for over 200,000 MeerLICHT images, we selected all trails away from the edges of the full-field images and matched them with catalogs obtained for the day of the observation from celestrak.org. This process involved cross-referencing the detected satellite trails with cataloged positions and trajectories using Two-Line Elements (TLEs), ensuring accurate identification of the satellites. We consider a detected trail to match a cataloged trail if the difference in their inclinations is less than 0.4 degrees, reflecting the high parallelism of well-matched trails. Additionally, the average distance between the ends of the two trails must be within a maximum of 200 arcseconds, accommodating minor shifts due to timing discrepancies between the known satellites and our exposure times.

\noindent
For this initial study, we concentrated on GEO and GES satellites, applying a declination cut between -15 and +25 degrees. Examining the distribution of the matched satellites, we selected all detected trails, whether positively or poorly matched, with trail lengths in pixels between 1440 and 1640, and oriented between 72.5 and 107.5 degrees, as there is a clear cutoff in the matched satellites at those degrees. Out of the remaining $9107$ detected trails, $7365$ ($80.9\%$) were matched to known satellites, while $1742$ ($19.1\%$) remained unmatched.
In Fig. \ref{fig:GEO_Detection}, we show both the matched and unmatched satellites. These form a thin band around the (projected) celestial equator. As seen from South Africa, part of the Virgo galaxy cluster lies projected behind the geostationary belt, leading to a large number of detections at RA$\sim$12hr, as these fields have been among the main targets of the MeerLICHT telescopes.

\begin{figure}[ht]
    \centering
    \includegraphics[width=\linewidth]{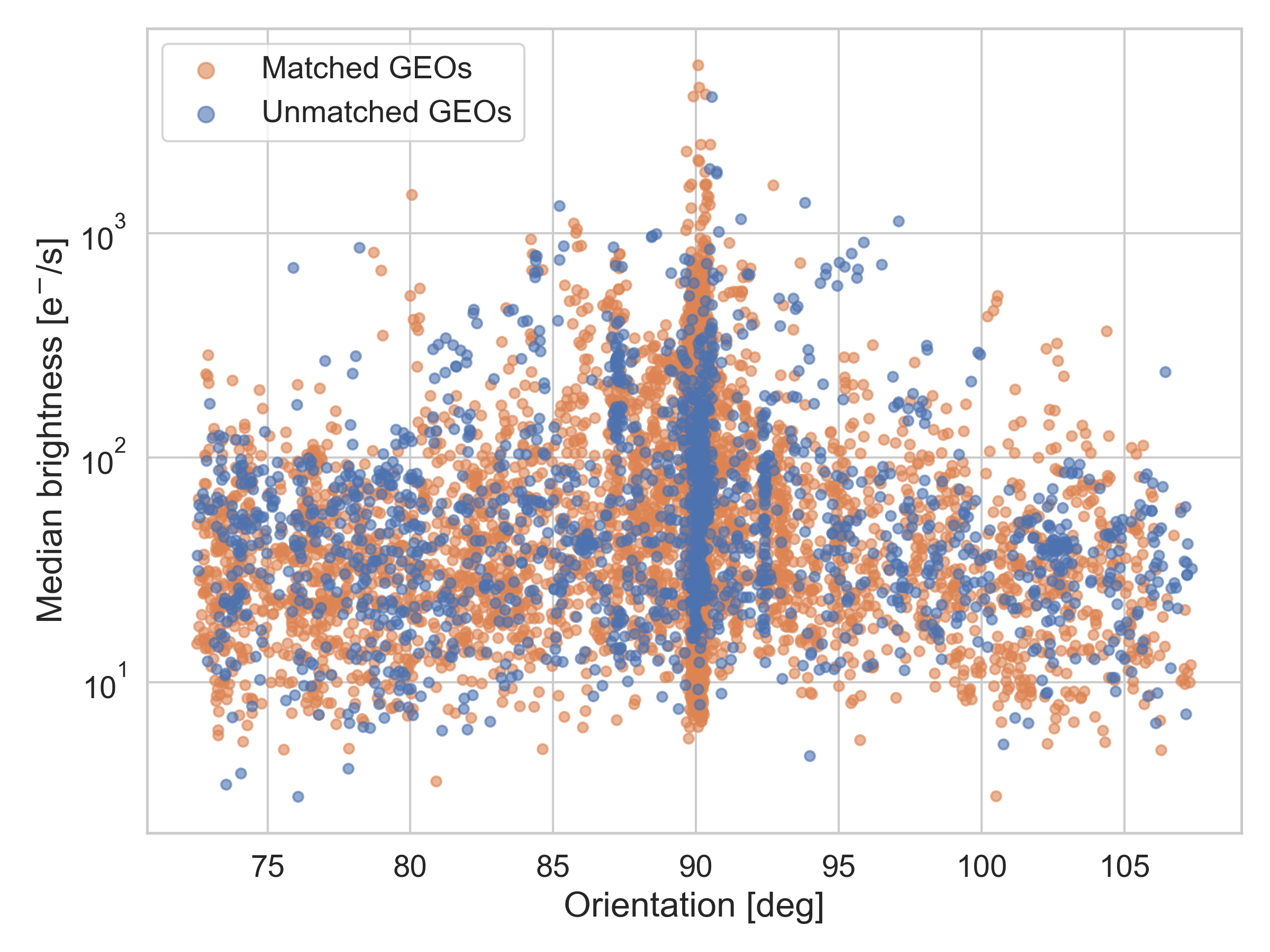}
    \caption{Comparison between matched and unmatched satellites in terms of orientation and median brightness of GEOs/GESs. $90\degree$ is E-W.}
    \label{fig:GEO_Bright}
\end{figure}

\begin{figure}[ht]
    \centering
    \includegraphics[width=\linewidth]{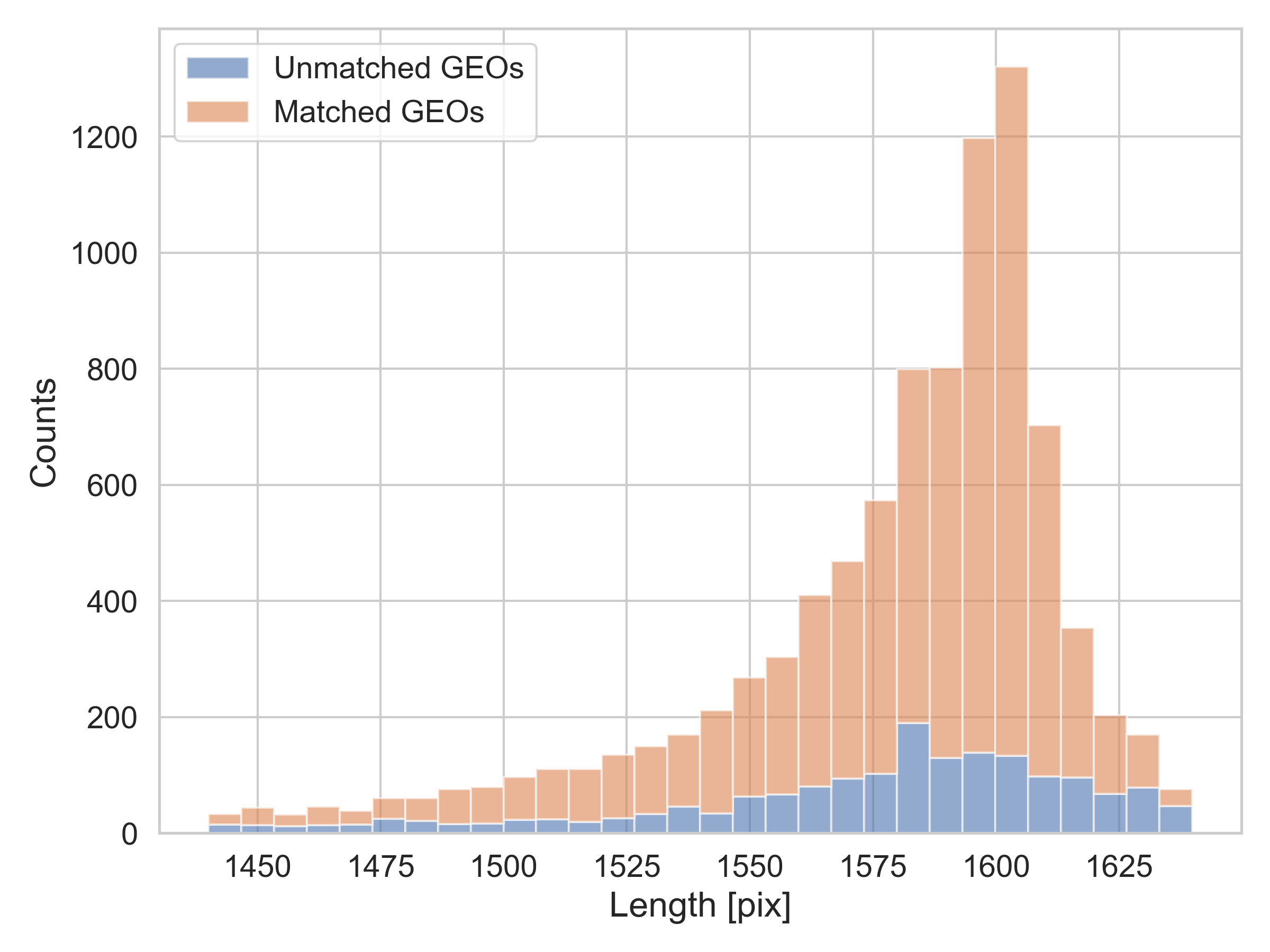}
    \caption{Stacked histogram of pixels length for matched and unmatched trails.}
    \label{fig:GEO_HistLength}
\end{figure}

\noindent
In addition to matched GEO and GES satellites, a substantial number of detections, $19.1\%$, cannot be matched to any object in public catalogues. A significant fraction of these unmatched satellites exhibit properties similar to those of the matched ones. Fig. \ref{fig:GEO_Bright} shows the comparison in terms of orientation and median brightness, while Fig. \ref{fig:GEO_HistLength} presents a stacked histogram of pixel lengths for both matched and unmatched trails, further supporting this observation.

\noindent
The similarity in properties between matched and unmatched satellites indicates the presence of a high fraction of publicly unknown (former) satellites or rocket stages. By identifying and cataloguing these unmatched satellites, especially if they are observed in multiple images taken within a short time frame and exhibit realistic orbital characteristics, we can contribute to the maintenance and expansion of public satellite catalogues. This effort is crucial for maintaining the safety and accuracy of future astronomical observations and could aid in the management and mitigation of space debris.


\section{Conclusions}
\label{sec: Conclusions}

In this study, we introduced ASTA (Automated Satellite Tracking for Astronomy), a robust methodology combining U-Net and Probabilistic Hough Transform to detect and analyze satellite trails in ground-based astronomical observations. Using data from the MeerLICHT telescope, we demonstrated the effectiveness of ASTA in identifying and characterizing satellite trails. 

\noindent
Importantly, the methodology developed in this study can be easily adopted by other observatories. The use of LABKIT for manual annotation ensures a straightforward and reproducible process, encouraging other researchers and observatories to implement similar techniques. By sharing this approach, we hope to foster a collaborative effort in addressing the challenges posed by satellite trails, improving data quality across various astronomical facilities.

The precision of ASTA in detecting satellite trails was validated against a comprehensive test set, showcasing high accuracy and robustness. The integration of deep learning and classical image processing techniques proved effective in refining trail detection, even in the presence of bright stars and other complex backgrounds.

\noindent
ASTA was applied to approximately 200,000 MeerLICHT images, and the results were cross-referenced with public satellite catalogs. We focused on identifying geostationary and geosynchronous satellites, successfully matching $81\%$ of the detected trails with known objects. Additionally, the discovery of satellites not found in public catalogs underscores gaps in current cataloging efforts and reveals the presence of previously uncatalogued objects.

Future improvements could enhance artefact recognition and overall model performance. One key improvement is the use of better and larger GPUs, which would eliminate the need for patching. This would provide U-Net with more context from larger images and nearby sources, reducing confusion caused by small patches and minimizing gaps or inconsistencies in predictions. By processing larger portions of the image at once, U-Net can maintain context and continuity across the entire field, leading to more accurate and reliable detections. Additionally, increased computational efficiency from advanced GPUs will allow for faster processing times, making it feasible to analyze large datasets of astronomical images more efficiently.

Future work will concentrate on a comprehensive statistical analysis of the temporal and spatial components of all types of satellites, from GEOs to LEO satellites, using MeerLICHT and BlackGEM data collected over the last five years. This analysis will examine trends and patterns, improving our understanding of satellite distributions and their impact on observational data, providing a basis for developing more effective mitigation strategies.

\vspace*{2\baselineskip} 

\begin{acknowledgements}
We thank Dr Marco Langbroek for the insightful discussions on satellite trails and the various observational biases associated with them.
PJG is partly supported by SARChI Grant 111692 from the South African National Research Foundation. MeerLICHT is designed, built and operated by a consortium of universities and institutes, consisting of Radboud University, the University of Cape Town, the South African Astronomical Observatory, the University of Oxford, the University of Manchester and the University of Amsterdam. 
\end{acknowledgements}

\bibliography{Bibliography}

\begin{appendix}

\section{Computation Time}
\label{sec: Computation Time}

To evaluate the efficiency of ASTA, we measured the computation time required to process a full-field MeerLICHT image, including image loading, patch creation, U-Net prediction, Hough Transform application, and satellite information extraction. The tests were conducted on an Alienware Area 51M equipped with an Intel Core i9-9900K processor, 32GB DDR4/2400 RAM, and an Nvidia GeForce RTX 2080 GPU.

The processing times were measured using both CPU and GPU for each of the four stages: Preprocessing, Prediction, Hough Transform, and Contour Analysis. The results are summarized in Table \ref{tab:times}.

\noindent
These results demonstrate the substantial efficiency gains achievable through GPU acceleration. The average total GPU processing time was significantly lower than the total CPU processing time, highlighting the critical role of GPU acceleration in handling large volumes of astronomical data effectively.

\begin{table}[htbp]
\centering
\caption{Average computation time and standard deviation for each processing stage using CPU and GPU.}
\begin{tabular}{l|c|c}
Stage & CPU Time (s) & GPU Time (s) \\
\hline
\hline
Preprocessing & $0.66 \pm 0.01$ & $0.66 \pm 0.01$ \\
Prediction & $21.97 \pm 0.25$ & $2.56 \pm 0.04$ \\
Hough Transform & $0.13 \pm 0.02$ & $0.13 \pm 0.02$ \\
Contour Analysis & $1.20 \pm 0.26$ & $1.24 \pm 0.26$ \\
\hline
Total & $23.96 \pm 0.53$ & $4.59 \pm 0.33$ \\
\end{tabular}
\label{tab:times}
\end{table}

\noindent
The Contour Analysis stage exhibits the highest variability in computation time, which can be attributed to the complexity and number of detected satellite trails in each image. The variability arises from the differing number of contours that need to be processed and the presence of intersecting trails, which require additional computation to separate accurately.

\end{appendix}

\end{document}